\documentclass{WileyMSP-template}

\usepackage{amssymb}
\usepackage{color} 

\begin{document}

\pagestyle{fancy}
\rhead{\includegraphics[width=2.5cm]{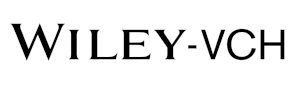}}

\title{The use of the correlated Debye model for EXAFS-based thermometry in bcc and fcc metals }

\maketitle


\author{Alexei Kuzmin*},
\author{Vitalijs Dimitrijevs},
\author{Inga Pudza},
\author{Aleksandr Kalinko}


\dedication{}

\begin{affiliations}
Dr.\ A.\ Kuzmin, Mg. V.\ Dimitrijevs, Dr.\ I.\ Pudza, Dr.\ A.\ Kalinko\\
Institute of Solid State Physics, University of Latvia, Kengaraga Street 8, LV-1063 Riga, Latvia   \\
Email Address: a.kuzmin@cfi.lu.lv
\end{affiliations}



\begin{abstract}
Extended X-ray absorption fine structure (EXAFS) spectra are sensitive to thermal disorder and are often used to probe local lattice dynamics. Variations in interatomic distances induced by atomic vibrations are  described by the temperature-dependent mean-square relative displacement (MSRD), also known as the Debye-Waller factor. In this study, we evaluated the feasibility of addressing the inverse problem, i.e., determining the sample temperature from the analysis of its EXAFS spectrum using the multiple-scattering formalism, considering contributions up to the 4th-7th coordination shell. The method was tested on several monatomic metals (bcc Cr, Mo, and W; fcc Cu and Ag), where the correlated Debye model of lattice dynamics provides a fairly accurate description of thermal disorder effects up to distant coordination shells. We found that the accuracy of the method  strongly depends on the temperature range. The method fails at low temperatures, where quantum effects dominate and MSRD values change only slightly. However, it becomes more accurate at higher temperatures, where the MSRD shows a near-linear dependence on temperature.
\end{abstract}

\section{Introduction}
\label{secintro}
\medskip
Temperature is a fundamental factor that influences physical, chemical, and biological phenomena. However, conventional methods of temperature measurement are not applicable at the nanoscale due to either the size restrictions of the thermometer or limited access to the region of interest \cite{Quintanilla2018}. To address these challenges, the field of nanothermometry has emerged, aiming to develop appropriate tools for local temperature measurement \cite{Quintanilla2018,Brites2012,Brites2019,Bradac2020,Zhou2020,Ye2024}. In addition to its practical importance, the development of nanothermometry offers an opportunity to gain a better understanding of the microscopic limits of thermodynamics \cite{Hartmann2004,Hartmann2006}. 

\medskip
The micro- and nanothermometry techniques can be grouped into three categories: (1) contact techniques, mainly based on scanning microscopes, (2) non-contact optical techniques, and (3) semi-contact techniques based on luminescent nanoparticles. The advantages and disadvantages of these techniques have been reviewed in detail in the literature (see, for example, \cite{Quintanilla2018,Brites2012}). Contact techniques are surface-limited, as the probe must be placed in contact with the desired point of the object under study and remain there until thermal equilibrium is reached. Non-contact techniques are based on various optical methods, allowing for remote data collection, often from the inside of the sample when it is transparent. However, their spatial resolution is limited by the wavelength of light. Additionally, some limitations are specific to the technique. For instance, the use of Raman spectroscopy is constrained by black-body radiation at high temperatures (above about 1000~K)  and the weak population of vibrational modes at low temperatures. Furthermore, the emission of fluorescent species present in the sample may obscure the Raman signal. Finally, most pure metals are not Raman-active. Semi-contact techniques, based on luminescent nanoparticles, are currently a very active area of research \cite{Quintanilla2018,Brites2019,Bradac2020,Kalaparthi2021,Peltek2022}. Similar to optical methods, their spatial resolution is diffraction-limited and  defined by the experimental optical setup. Additionally,  significant limitations  arise from variations in luminescence parameters (intensity, lifetime, and spectral shift) with particle concentration, fluctuations in excitation power, and environmental factors. 

\medskip
Extended X-ray absorption fine structure (EXAFS) spectra $\chi(k)$\  often exhibit strong temperature dependence due to changes  in interatomic distances caused by  atomic vibrations. This effect is accounted for by the EXAFS Debye-Waller term $\exp(-2 k^2 \sigma^2(T) )$, where $\sigma^2(T)$\ is  the temperature-dependent mean-square relative displacement (MSRD) or vibrational amplitude, and $k$ is the photoelectron wave number ($k = \sqrt{(2m_e/\hbar^2)(E-E_0)}$, where $m_e$ is the electron mass,  $\hbar$ is the reduced Planck constant, and $E_0$ is the threshold energy) \cite{Beni1976,Lee1981,Dalba1997}. 
The MSRD value $\sigma^2(T)$\ depends on the atomic mass and the strength of interatomic interactions and is often described using the Debye or Einstein  phenomenological models \cite{Beni1976,Bohmer1979,Sevillano1979}.  Therefore,  solving the inverse problem, i.e., determining temperature from the MSRD value, can be employed in non-contact nanothermometry.  Moreover, the element selectivity of EXAFS spectroscopy and the X-ray focusing capabilities provide significant opportunities for temperature mapping in compositionally and structurally non-homogeneous or nano-sized materials.

\medskip
It was proposed in \cite{VandeBroek2011} that the sensitivity of MSRD to thermal disorder could be used to determine the temperature  of plasmonic branched gold nanoparticles, approximately 54~nm in size, suspended in water under variable resonant laser illumination.  The authors first analyzed the Au L$_3$-edge EXAFS contribution from the first coordination shell of bulk gold to extract information on the MSRD temperature dependence in the range of 22--75.6~$^\circ$C, which was used for calibration \cite{VandeBroek2011}. A maximum temperature change of about 17~$^\circ$C was  detected under illumination with a He--Ne laser (633~nm) at a continuous wave (CW) power of 24.9~mW (3.2~W~cm$^{-2}$), and the accuracy of the temperature determination for gold nanoparticles was estimated to be approximately 6\% (or about 1~$^\circ$C) \cite{VandeBroek2011}.  

\medskip
Recently, a similar approach was used to determine the temperature of gold nanorods, approximately $41 \times 11$~nm in size, with the localized surface plasmon resonance (LSPR) located close to 800~nm, and Au/Fe hybrid star-like branched nanoparticles, approximately 35~nm in diameter,  with a broad LSPR peaked at about 650~nm \cite{Espinosa2021}.  Due to the nanometer size, the contribution from the distant coordination shells to the Au L$_3$-edge EXAFS spectrum is reduced. Therefore, only the first coordination shell was analyzed in  \cite{Espinosa2021} to extract information on the MSRD  temperature dependence in the range of 10--350~K and also after heating by a NIR laser (808~nm) with a controlled power between 0.05 to 0.6~W.  The uncertainty of the  procedure for temperature determination was reported to be below $\pm 5^\circ$  \cite{Espinosa2021}. A similar approach was also recently used  to determine the temperature of gold-iron oxide nanohybrids for photothermal  applications \cite{Lopez2023}.   

\medskip
In \cite{Ye2020}, the authors employed several approaches based on EXAFS cumulant expansion analysis and the anharmonic correlated Debye model to compare the experimental and determined temperatures for Ti, V, Fe, and Au metals in the high-temperature range of 300 to 700--1100~K.  The proposed approach resulted in temperature determination uncertainties of about 5--8\% at 300~K and about 16\% at 1000~K \cite{Ye2020}. 
 
\medskip
The temperature of heterodimer Fe$_3$O$_4$ nanocrystals with epitaxially attached Pt nanoparticles  was determined in \cite{Rosen2022} using the correlated Debye model for the MSRD temperature dependence obtained from the Pt L$_3$-edge.
The temperature uncertainty was about $\pm10$-15$^\circ$.   Pure Pt and Pt-Fe$_3$O$_4$ nanoparticles were either heated conventionally in the temperature range of 20--370~$^\circ$C or  inductively heated up to about 109~$^\circ$C in magnetic fields of  2--16.8~mT \cite{Rosen2022}.  A significant difference (up to $73.60 \pm 14.50$~$^\circ$C) between the temperatures of the nanocrystals, obtained from EXAFS, and their local support environment, determined by thermal imaging, demonstrates the capabilities  of  EXAFS-based nanothermometry  \cite{Rosen2022}.

\medskip
In this study, we used a set of reference metal foils with  body-centered cubic (bcc) (Cr, Mo, W) and face-centered cubic (fcc) (Cu, Ag) crystallographic structures to evaluate the uncertainty of temperature determination  from EXAFS spectra based on the multiple-scattering (MS) formalism and the correlated Debye model in the temperature range of 10--300~K.  
The use of the MS approach allowed us to extend the analysis beyond the first coordination shell, up to the 4th-7th shell.

\section{Experimental Section}

\subsection{Materials}

\medskip
High-purity commercial metal foils (Goodfellow Cambridge Ltd.)  of bcc Cr ($d$ = 5~$\mu$m, 99.99\%), fcc Cu ($d$ = 5~$\mu$m, 99.97\%), bcc Mo ($d$ = 5~$\mu$m, 99.99\%), fcc Ag ($d$ = 20~$\mu$m, 99.97\%), and bcc W ($d$ = 4~$\mu$m, 99.95\%) were used.  Their thicknesses, $d$, were optimized for X-ray absorption experiments.

\subsection{X-ray absorption experiments}
\label{secxas}

\begin{figure}
	\centering
	\includegraphics[width=0.5\linewidth]{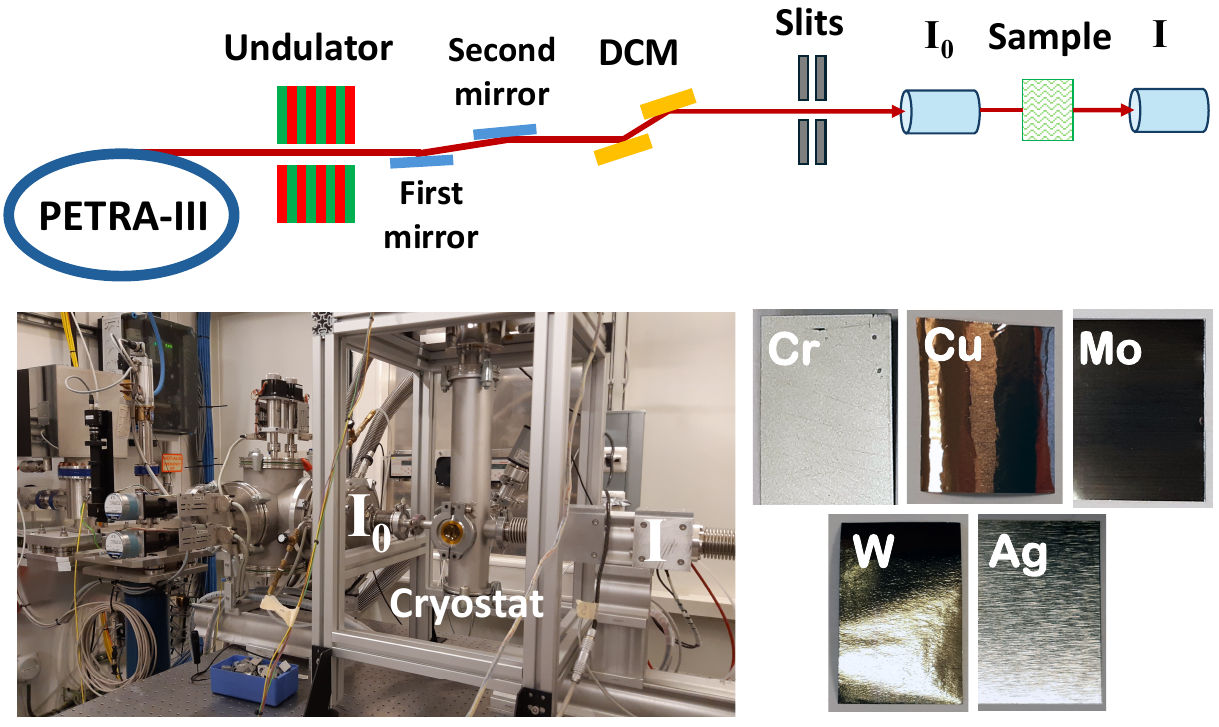}
	\caption{Schematics and a photo of the experimental setup at the DESY PETRA-III P65 Applied XAFS beamline, along with a set of metal foils used in the experiments. See the text for details. }
	\label{fig1}
\end{figure}

\medskip
Temperature-dependent X-ray absorption spectra of metal foils were recorded at the DESY PETRA-III P65 Applied XAFS beamline \cite{Welter2019} (Figure\ \ref{fig1}). The storage ring operated at an energy of $E = 6.08$~GeV and a current of $I = 120$~mA in top-up mode. The spectra were acquired in transmission detection mode using a continuous scan at the Cr (5989~eV), Cu (8979~eV), Mo (20000~eV), and Ag (25514~eV) K-edges, as well as  the W (10204~eV) L$_3$-edge. The X-rays from an undulator source were monochromatized using fixed-exit double-crystal monochromators (DCMs) Si(111) or Si(311). Harmonic reduction was achieved using two uncoated silicon plane mirrors (for Cr and Cu K-edges and W L$_3$-edge), Rh-coated mirrors (for Mo K-edge), or Pt-coated mirrors (for Ag K-edge). The X-ray spot size at the sample was 0.5$\times$1~mm$^2$.
The X-ray intensities before ($I_0$) and after ($I$) the sample were measured using two ionization chambers filled with nitrogen, argon, or krypton gases. The gas pressure in the chambers was optimized for each energy range. The X-ray absorption coefficient was calculated as $\mu_x(E) = ln[ I_0(E) / I(E) ]$.  
The selected thickness of the commercially available metallic foils resulted in absorption edge jumps $\Delta \mu_x$ of 1.5 for Cr, 1.1 for Cu,  1.1 for Mo, 1.0 for Ag, and 0.9 for W. A liquid helium flow cryostat (Janis Research Company, LLC) was used to maintain the required sample temperature in the range of 10–300~K. For each sample, the spectra were collected at seven or ten  temperatures. 

\subsection{Data analysis and simulations}
\label{secanalysis}

\medskip
The experimental extended X-ray absorption fine structure (EXAFS) spectra $\chi(k)$\  above each absorption edge were extracted using the XAESA code \cite{XAESA}, following a conventional procedure \cite{Kuzmin2014}. 
The Fourier transforms (FTs) of the EXAFS spectra were calculated using a  10\% Gaussian function.  It is important to note that the FTs of the EXAFS spectra were not corrected for the phase shift present in the EXAFS equation. As a result, the positions of the peaks in the FTs are shifted to shorter distances compared to their crystallographic values.

\medskip
The simulations of temperature-dependent EXAFS spectra were performed using the ab initio multiple-scattering formalism \cite{Rehr2000}, taking into account thermal disorder effects according to the correlated Debye model within the harmonic approximation  \cite{Beni1976,Bohmer1979,Sevillano1979,Scardi2018}.  The total EXAFS spectrum  $\chi(k, T)$\ was calculated as a sum over all multiple-scattering contributions ($N_{\rm tot}$) with the scattering path lengths $2R_i$ up to a selected maximum value $2R_{\rm max}$ (Table\ \ref{table1}):
\begin{equation}
 \chi(k, T) = S_0^2 \sum_{i=1}^{N_{\rm tot}} \chi_i(k, R_i(T), Amp_i(k), Pha_i(k))) \exp(-2 \sigma_i^2(T) k^2 ).
 \label{eq1}
\end{equation}
where  $\sigma_i^2(T)$\ is the temperature-dependent mean-square relative displacement (MSRD) or Debye-Waller factor. The $R_{\rm max}$ value was selected for each metal to include scattering paths that contribute to about 6~\AA\ in the FTs of the respective EXAFS spectra. This choice includes multiple-scattering contributions up to the 4th order, i.e., single-scattering, double-scattering, and triple-scattering events. The total number of scattering paths $N_{\rm tot}$ depends on the type of  lattice (bcc or fcc) and the lattice constant ($a_0$), as reported in Table\ \ref{table1}. The scattering path degeneracies, equal to the coordination numbers for the single-scattering paths, were determined by the crystal lattice symmetry.

\begin{table}
	\caption{Parameters used in the EXAFS spectra calculations.
		$a_0$\ is the lattice constant \protect\cite{Straumanis1955,Jette1935,Dubrovinsky1997,Owen1933,Suh1988}, $R_{\rm max}$\ is the length of the maximal scattering path included in the calculations, $N_{\rm tot}$\ is the total number of scattering paths.  }
	\label{table1}	
	\begin{tabular}[htbp]{@{}llll@{}}
		\hline
		Metal &   $a_0$\ [\AA] & $R_{\rm max}$\ [\AA] & $N_{\rm tot}$\   \\
		\hline
		bcc Cr & 2.880 & 6.277 & 82 \\                 
		bcc Mo & 3.141 & 6.186 & 37 \\
		bcc W  & 3.165 & 5.897 & 34 \\   	   
		&           &            &      \\    
		fcc Cu & 3.610  &  6.253 &  59  \\     
		fcc Ag & 4.090  &  5.996  &  24 \\      
		\hline
	\end{tabular}
\end{table}

\medskip
The scattering amplitude $Amp_i(k)$\ and phase shift $Pha_i(k)$\ functions for the required multiple-scattering paths were calculated for ideal crystallographic structures of metals using the ab initio self-consistent real-space MS FEFF8.5L code \cite{Ankudinov1998}. 
Inelastic effects were accounted for using the complex energy-dependent exchange-correlation Hedin-Lundqvist potential \cite{Hedin1971}.

\medskip
The temperature dependence $\sigma_i^2(T)$\ can be calculated in the harmonic approximation as \cite{Sevillano1979,Rehr2000} 
\begin{equation}
	\sigma_i^2(T) = \frac{\hbar}{2 \mu_i} \int_0^{\omega_{\rm max}}  \frac{\rho_i(\omega)}{\omega} \coth \left(  \frac{\hbar \omega }{2 k_B T} \right)  d\omega 
\end{equation}
where $\mu_i$\  and  $\rho_i(\omega)$\  are, respectively, the reduced mass and the projected local vibrational density of states for the scattering path $i$,  $k_B$\ is the Boltzmann constant, and $\omega_{\rm max}$\ is the maximum frequency. In this study, we used the correlated Debye model \cite{Beni1976,Bohmer1979,Sevillano1979}, which is often applied to metals and provides reasonable accuracy for lattice dynamics in the case of bcc or fcc lattices \cite{Jeong2003}. 
The model accounts for the phonon dispersion effect and the acoustic phonon branch in crystals. 
In this case, the projected vibrational density of states for a pair of atoms located at a distance $R$ is given by  \cite{Sevillano1979}
\begin{equation}
 \rho_R(\omega) = \frac{3 \omega^2}{\omega_{\rm D}^3}  \left[ 1 - \frac{\sin( \omega R / c) }{ \omega R / c }  \right] 
\end{equation}
where  $\omega_{\rm D} = k_B \theta_{\rm D} / \hbar$\ is the Debye cut-off frequency, and $\theta_{\rm D}$\ is the Debye temperature. $c = \omega_{\rm D} / k_{\rm D}$\ is the Debye approximation for the speed of sound in a material with  atomic number density $n$\ and $k_{\rm D} = (6 \pi^2 n)^{1/3}$. 
It is convenient for practical applications that the Debye model for a particular material depends solely on a single parameter ($\theta_{\rm D}$).

\medskip
In the first round, the fit of each set of temperature-dependent EXAFS spectra $\chi(k)k^2$\ was performed using the FEFFIT code \cite{Newville1995,Newville2001} in the back-transformed $k$-space from 2.5 to 17.0~\AA$^{-1}$. 
This approach allows for simultaneous analysis of the experimental EXAFS spectra  recorded at several temperatures using a strongly limited number of fitting parameters. Indeed, our EXAFS model included only five variable parameters: $S_0^2$, $\Delta E_0$, $\alpha_0$, $\alpha_{\rm T}$, and $\theta_{\rm D}$. The first two non-structural parameters are  responsible for correcting the EXAFS  amplitude ($S_0^2$) and the photoelectron energy origin ($\Delta E_0$). 
The constant lattice expansion $\alpha_0$ and the thermal expansion coefficient $\alpha_{\rm T}$\ were used to account for  lattice expansion.  As a result of the fits, the values of the Debye temperatures, $\theta_{\rm D}$, reported in the third column of Table\ \ref{table2}, were obtained.  The MSRD values obtained from the correlated Debye model can be found in the Supporting Information (see Tables S1-S5).

\begin{table}
	\caption{Parameters of the correlated Debye model \protect\cite{Beni1976,Bohmer1979,Sevillano1979} for five metals with bcc and fcc crystallographic structures.  Values of the Debye temperature $\theta_{\rm D}$ from the literature \protect\cite{Ho1972,Pirog2002,Pirog2003,Hung2010,Vanhung2016,Tien2024} are shown in the second column, while those determined in the present study are given in the third column. $T_{\rm calc}$ is the temperature obtained from the fit of the experimental EXAFS spectrum measured at the temperature $T_{\rm exp}$.}
	\label{table2}	 
	\begin{tabular}[htbp]{@{}lllllll@{}}
		\hline
		Metal & $\theta_{\rm D}$\ [K]  & $\theta_{\rm D}$\ [K] & $T_{\rm calc}$ [K] & $T_{\rm calc}$ [K] &  $T_{\rm calc}$ [K] &  Number of variable parameters \\
	          &  &  this study    & ($T_{\rm exp}$ = 10--16~K) & ($T_{\rm exp}$ = 150~K) &  ($T_{\rm exp}$ = 293--300~K) &   \\		
		\hline
		bcc Cr & 424 \protect\cite{Ho1972} & 412$\pm$16 & 151$\pm$34 &  183$\pm$32 &  272$\pm$34  &  5 \\                
			   & 427 \protect\cite{Pirog2003}					&            &  93$\pm$21 &  128$\pm$17 &  222$\pm$16  &  1 \\ 
		& &           &            &           & 	           &   \\  	
		bcc Mo & 377 \protect\cite{Ho1972} & 408$\pm$5 & 71$\pm$29&  148$\pm$16 &  296$\pm$18  &  5 \\
				& 374.5$\pm$17 \protect\cite{Pirog2003} 					&           & 53$\pm$24  & 146$\pm$8 & 	304$\pm$9  &  1 \\
		& 395.3  \protect\cite{Vanhung2016} &           &            &           & 	           &   \\   				
		& &           &            &           & 	           &   \\   
		bcc W & 312 \protect\cite{Ho1972} & 333$\pm$4  &  134$\pm$16 &  195$\pm$14 &  314$\pm$12  &  5 \\   	   
				& 302.6$\pm$17  \protect\cite{Pirog2003}					&           &  67$\pm$11  &   149$\pm$6 & 286$\pm$6    &  1 \\    
		& 296.9 \protect\cite{Vanhung2016} &           &            &           & 	           &   \\  					
		& &           &            &           & 	           &   \\            
		fcc Cu & 310 \protect\cite{Ho1972} & 321$\pm$3  &  55$\pm$22 &  152$\pm$15 &  314$\pm$23  &  5 \\ 
				&301$\pm$22 \protect\cite{Pirog2002}	&            &  43$\pm$15  &  148$\pm$7 &  293$\pm$9  &  1 \\       
		&333 \protect\cite{Hung2010}  &           &            &           & 	           &   \\    				
		& &           &            &           & 	           &   \\           
		fcc Ag & 221 \protect\cite{Ho1972} & 222$\pm$2  &  54$\pm$8 &  153$\pm$9&  307$\pm$21  &  5 \\      
				& 231.7 \protect\cite{Tien2024}	&            &  38$\pm$6  &  143$\pm$4 &  283$\pm$8  &  1 \\     
		&245.7 \protect\cite{Tien2024}  &           &            &           & 	           &   \\				
		\hline
	\end{tabular}
\end{table}

\medskip
In the second round, the Debye temperatures, $\theta_{\rm D}$, obtained were always fixed, while two models were used to evaluate the temperatures from the three selected experimental EXAFS spectra, which were measured at low (10--16~K), intermediate (150~K), and high (293~K or 300~K) temperatures $T_{\rm exp}$. In the first model, five variable parameters ($S_0^2$, $\Delta E_0$, $\alpha_0$, $\alpha_{\rm T}$, and $T_{\rm calc}$) were employed (see the last column in Table\ \ref{table2}). In the second model, only the temperature  ($T_{\rm calc}$) was varied, while the other parameters were fixed at the values obtained during the first round.  The obtained temperature values $T_{\rm calc}$ and their fitting uncertainties are reported in Table\ \ref{table2}.

\section{Results and Discussion}
\label{secresults}

\begin{figure}
	\centering
	\includegraphics[width=0.5\linewidth]{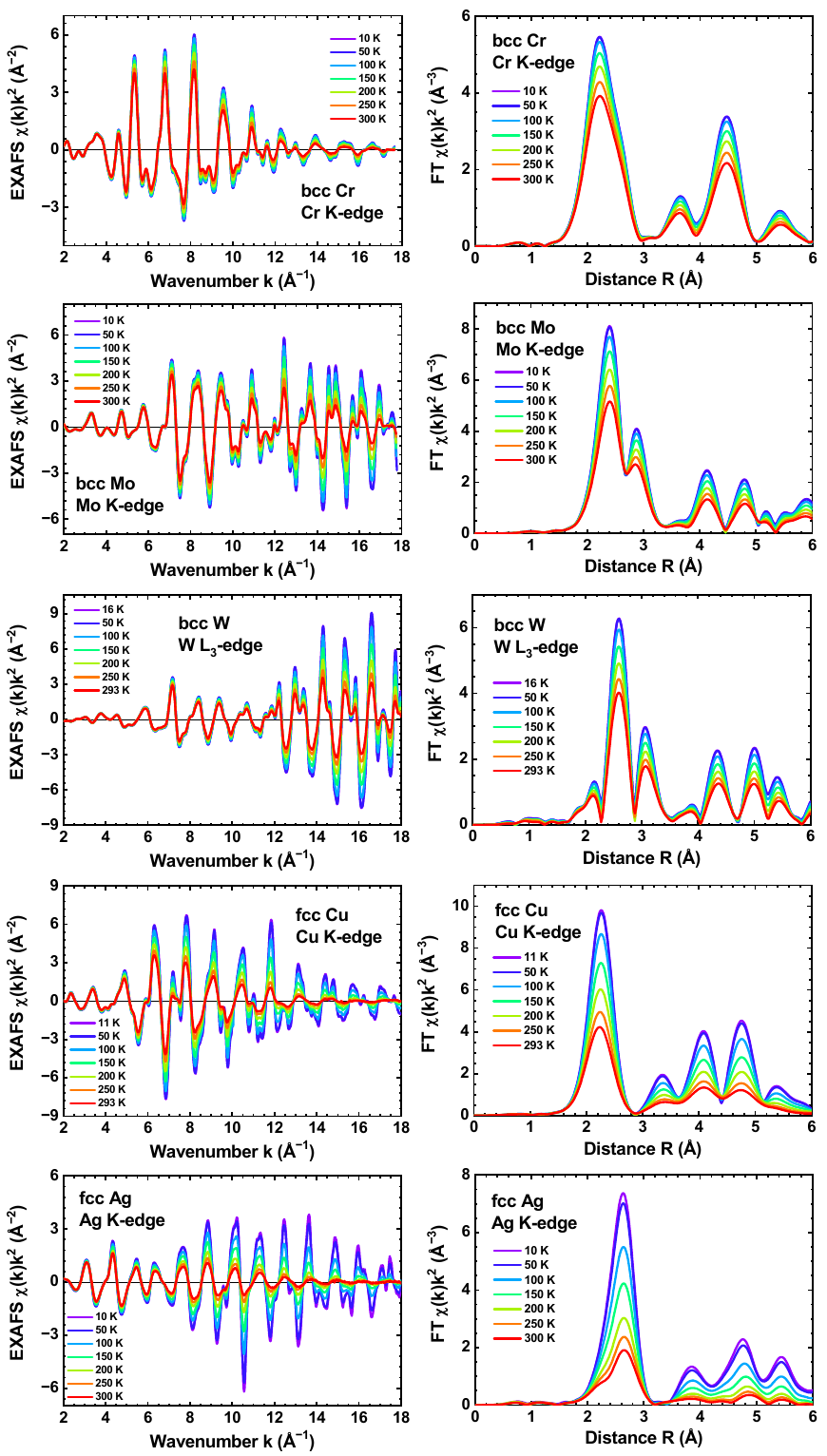}
	\caption{Temperature-dependent K-edge EXAFS spectra $\chi(k)k^2$\ and their Fourier transforms (FTs) for bcc and fcc metals. }
	\label{fig2}
\end{figure}

\medskip
Experimental  EXAFS spectra $\chi(k)k^2$\ and their FTs for bcc Cr, Mo, and W, as well as for fcc Cu and Ag, are shown in \textbf{Figure\ \ref{fig2}} as a function of temperature in the range of 10--300~K. Note that the use of metal foils with optimized thicknesses resulted in experimental data with a high signal-to-noise ratio across a wide $k$-space range up to 18~\AA$^{-1}$.  An increase in temperature from 10 to 300~K causes a  stronger damping of the EXAFS spectrum amplitude at larger  $k$ values due to the exponential Debye-Waller term,  and a decrease in the FT peak intensity while leaving the peak positions nearly unaffected. This  indicates that  the thermal disorder effects related to a change in the MSRD $\sigma^2(T)$\ are dominant.  Additionally, one can observe the effect of the scattering amplitude ($Amp(k)$\ in Equation\ (\ref{eq1})) on the overall shape of the EXAFS spectra $\chi(k)k^2$: the envelope maximum shifts from about $k = 8$~\AA$^{-1}$ for light chromium to $k = 16$~\AA$^{-1}$\ for heavy tungsten. 
Note that the EXAFS spectra do not show any visible phase shift (except for the Ag foil) because the lattice expansion of the studied metals is small in the  temperature range of 10-300~K. At the same time, for the Ag foil,  the phase shift at high $k$-values is masked due to large MSRD values, leading to strong damping of the Ag K-edge EXAFS amplitude.

\medskip
The difference between bcc (Cr, Mo, W) and fcc (Cu, Ag) structures is clearly observed in the shape of the respective FTs (Figure\ \ref{fig2}). In the bcc structure, the first peak in the FT is distorted for Cr and even split for Mo and W because it includes contributions  from the first and second coordination shells, containing  8 and 6 atoms, respectively. In contrast, in the fcc structure, there is one group of 12 atoms in the first coordination shell, and, therefore, the first peak in the FT is more symmetric. Note that the nonlinearity in the scattering phase $Pha(k)$\ in Equation\ (\ref{eq1}) \cite{Lee1981} also affects the shape of the peaks in the FTs.

\medskip
Depending on the lattice parameter $a_0$\ (Table\ \ref{table1}) and the type of lattice (bcc or fcc),  clusters with varying sizes $R_{\rm max}$, including 5 coordination shells (for Mo, W, and Ag), 6  coordination shells (for Cu), and 7 coordination shells (for Cr)  around the absorbing atom, are required to describe the EXAFS contributions that give rise to the peaks in the FTs up to about 6~\AA.  Therefore, the total number  $N_{\rm tot}$\  of scattering paths was different for each metal. 

\begin{figure}
	\centering
	\includegraphics[width=0.5\linewidth]{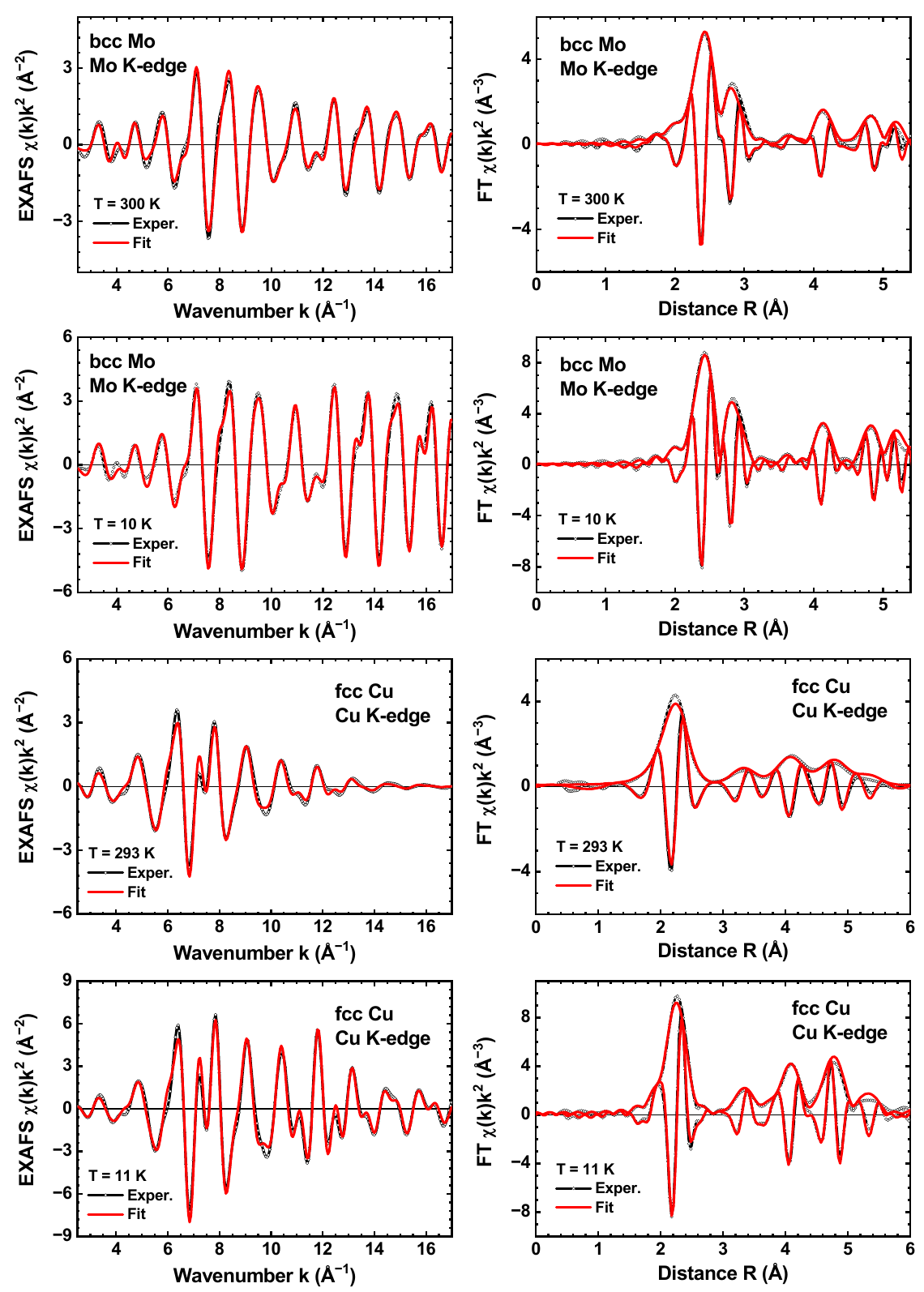}
	\caption{Examples of the experimental (open circles) and calculated (solid lines)  K-edge EXAFS spectra $\chi(k)k^2$\ and their Fourier transforms (FTs) for bcc molybdenum at 10~K and 300~K and fcc copper at 11~K and 293~K.  }
	\label{fig3}
\end{figure}

\medskip
To determine the characteristic Debye temperatures, $\theta_{\rm D}$, the set of temperature-dependent EXAFS spectra $\chi(k)k^2$\ for each metal was simultaneously fitted using the FEFFIT code \cite{Newville1995,Newville2001}, taking into account the  $N_{\rm tot}$\ scattering paths up to the 4th order, as described in Section\ \ref{secanalysis}. Selected examples of fits for bcc Mo and fcc Cu at low and high temperatures are shown in \textbf{Figure\ \ref{fig3}},  demonstrating good agreement between the experimental and calculated EXAFS spectra in both $k$ and $R$ space. Similar fit quality was achieved for other metals and temperatures.  The obtained values of the Debye temperatures, $\theta_{\rm D}$,  are reported in the third column in Table\ \ref{table2}. They are in reasonable agreement with the literature data from \cite{Ho1972,Pirog2002,Pirog2003,Hung2010,Vanhung2016,Tien2024}, which are reported in the second column. The deviation between the Debye temperatures obtained in different studies ranges from 
3~K for bcc Cr \cite{Ho1972,Pirog2003} to 32~K for fcc Cu \cite{Pirog2002,Vanhung2016}. 
Note that in previous studies, analysis of the first peak in the FTs of EXAFS spectra was used to evaluate the anharmonicity of nearest-neighbor interactions, typically based on anharmonic pair potentials or cumulant expansion. 
Anharmonic effects are important for describing phenomena such as thermal lattice expansion and high-temperature lattice dynamics. Nevertheless, they represent small corrections to the MSRD values, especially in the outer coordination shells analyzed here. Therefore, anharmonicity corrections were neglected in the present study.

\begin{figure}
	\centering
	\includegraphics[width=0.5\linewidth]{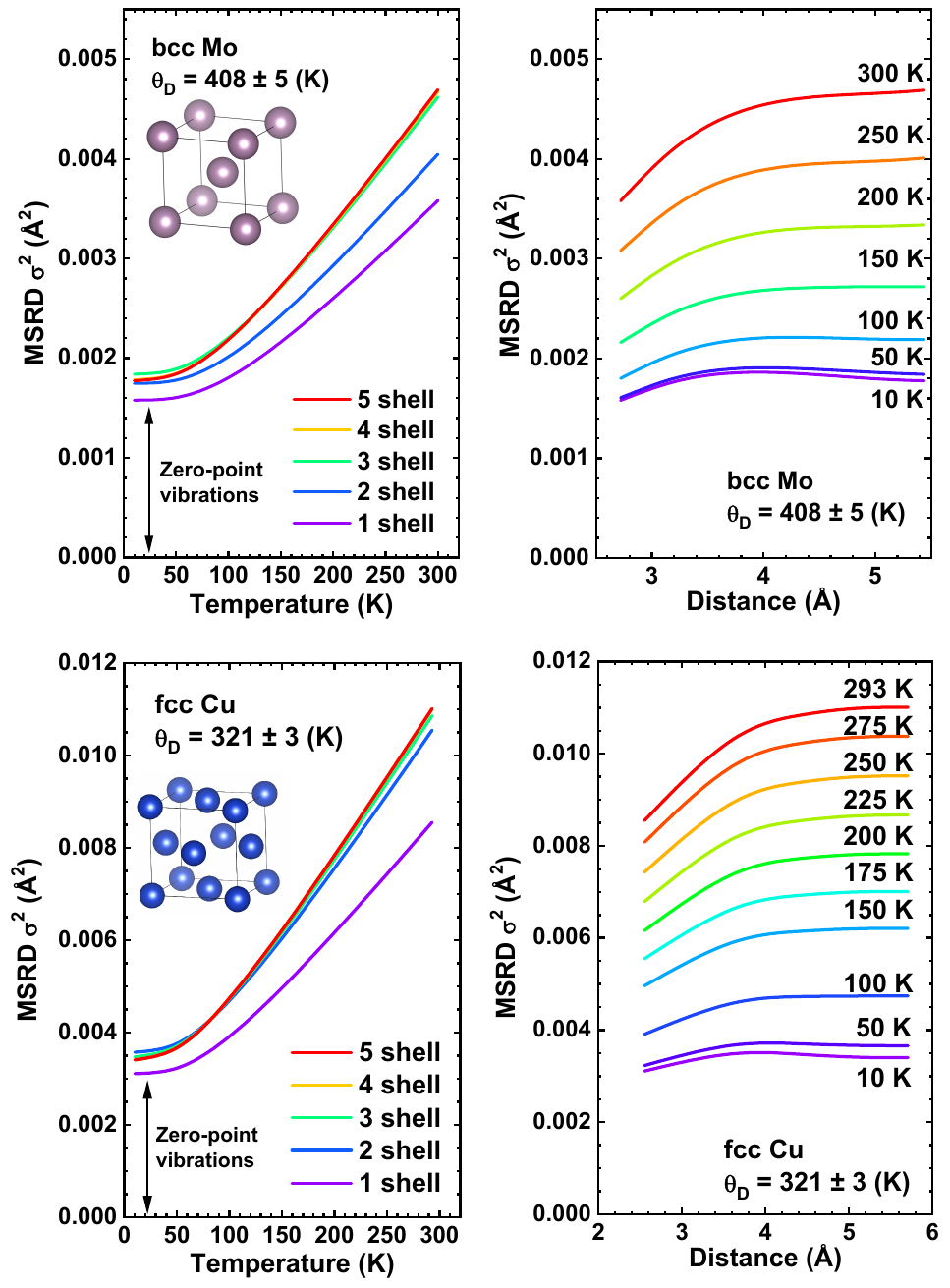}
	\caption{Dependence of the mean-square relative displacement (MSRD) $\sigma^2$\  on temperature and interatomic distance for the first five coordination shells in bcc molybdenum and fcc copper.  $\theta_{\rm D}$\ is the Debye temperature. The contribution of  zero-point vibrations to MSRD is indicated.}
	\label{fig4}
\end{figure}

\medskip
The MSRD values $\sigma^2(T,R)$, calculated for bcc Mo  and fcc Cu  using the correlated Debye model within the harmonic approximation  \cite{Beni1976,Bohmer1979,Sevillano1979}
and  the Debye temperatures, $\theta_{\rm D}$, (Table\ \ref{table2}),  are shown in \textbf{Figure\ \ref{fig4}} as a function of temperature and interatomic distance for the first five coordination shells. The obtained dependencies allow for three conclusions. First, at low temperatures, the MSRD depends weakly on temperature, and its value is determined by zero-point motion. Indeed, it was  demonstrated in \cite{Yang2012} that within the Debye model, the magnitude of zero-point vibrations equals that of the excited vibrations at $T_0 \sim 1/3\theta_{\rm D}$.
Second, at temperatures above $T_0$, the temperature dependencies of the MSRDs are almost linear, which is convenient for use in thermometric applications. The difference between the MSRD values  for different coordination shells is due to the change in the correlation of atomic motion  \cite{Jeong2003} and is clearly visible in Figure\ \ref{fig4}. The correlation decreases with decreasing temperature and increasing interatomic distance, i.e., for distant coordination shells \cite{Bohmer1979}. 
Third, the dependence of MSRD on interatomic distance suggests that the MSRD value approaches some limit at large distances due to the progressive loss of correlation in atomic motion. As a result, the MSRD for a pair of atoms becomes equal to the sum of the uncorrelated mean-square displacements (MSDs) of these atoms \cite{Jeong2003}. Therefore, the analysis of MSRD values for  distant coordination shells allows for the determination of MSD values, as demonstrated for tungsten in \cite{Jonane2018}. 
Note also that, since the Debye model assumes an isotropic crystal structure, it is less accurate for the nearest coordination shells where correlation effects are important \cite{Jeong2003}.

\medskip
To check the accuracy of using the correlated Debye model for EXAFS-based thermometry in bcc and fcc metals, we employed two models (Section\ \ref{secanalysis})  with five variable parameters ($S_0^2$, $\Delta E_0$, $\alpha_0$, $\alpha_{\rm T}$, and $T_{\rm calc}$)  and  one variable parameter ($T_{\rm calc}$), respectively.  The experimental EXAFS spectra $\chi(k)k^2$,  measured at three selected temperatures $T_{\rm exp}$ -- low (10--16~K), intermediate (150~K), and high (293~K or 300~K) -- were fitted. These temperatures were chosen to test the accuracy of the approach  in regions where zero-point vibrations dominate, close to the beginning of the  MSRD $\sigma^2(T)$\  linear slope, and at the  MSRD $\sigma^2(T)$\ linear slope, i.e., in the classical limit. The obtained temperatures $T_{\rm calc}$ and their uncertainties  are reported in Table\ \ref{table2}. 
As might be expected, both models fail at low temperatures, where the MSRD has weak temperature dependence, making the solution unstable.  At intermediate and high temperatures, the temperature values obtained by the two models are close within the  error margins. The average difference between the experimental and obtained temperatures is less than 10~K for all metals except chromium, where it is about four times larger. 
While the involvement of several coordination shells and accounting for multiple-scattering contributions play a positive role in the analysis, the intrinsic limitations of the Debye model  affect  the accuracy of the method.

\section{Conclusion}
\label{secconc}

\medskip
The sensitivity of the EXAFS amplitude to thermal disorder provides an opportunity to determine the sample temperature by analyzing variations in interatomic distances induced by atomic vibrations \cite{VandeBroek2011,Espinosa2021,Ye2020,Rosen2022,Ano2020}.  In this study, we tested this idea using several monatomic metals with bcc (Cr, Mo, and W) and fcc (Cu and Ag) structures over a temperature range of 10--300~K in harmonic approximation.  

\medskip
The temperature dependencies of MSRDs, also known as the EXAFS Debye-Waller factors, for the nearest 4--7 coordination shells were approximated using the correlated Debye model \cite{Beni1976,Bohmer1979,Sevillano1979}, which was employed in the analysis of EXAFS spectra based on the multiple-scattering formalism \cite{Rehr2000}. The dependence of the Debye model on a single parameter, $\theta_{\rm D}$, is an advantage, though it is limited by a poor description of the phonon density of states.   

\medskip
The obtained Debye temperatures  were  used to predict the sample temperature. It was found that the accuracy of the method  depends strongly on the temperature range. In particular, EXAFS-based thermometry fails  at low temperatures, where quantum effects (zero-point vibrations) dominate and MSRD values change weakly with temperature. At high temperatures, where the MSRD dependence on temperature is nearly linear, a more accurate estimate of the sample temperature is possible.



\medskip
\textbf{Acknowledgements} \par 
This study was supported by the Latvian Council of Science project No. LZP-2022/1-0608.
We acknowledge DESY (Hamburg, Germany), a member of the Helmholtz Association HGF, for the provision of experimental facilities. 
Parts of this research were carried out at PETRA III and we would like to thank Dr. Edmund Welter for his assistance in using the P65 beamline. 
Beamtime was allocated for the proposal I-20220209 EC. 
The Institute of Solid State Physics, University of Latvia as the Center of Excellence has
received funding from the European Union’s Horizon 2020 Framework Programme H2020-WIDESPREA-D-01-2016-2017-TeamingPhase2 under grant agreement No. 739508, project CAMART2.

\medskip
\textbf{Conﬂict of Interest} 
The authors declare no conﬂict of interest.

\medskip
\textbf{Data Availability Statement}
The data that support the ﬁndings of this study are available from the corresponding author upon reasonable request.

\medskip

%

\newpage

\begin{figure}
\textbf{Table of Contents}\\
\medskip
  \includegraphics{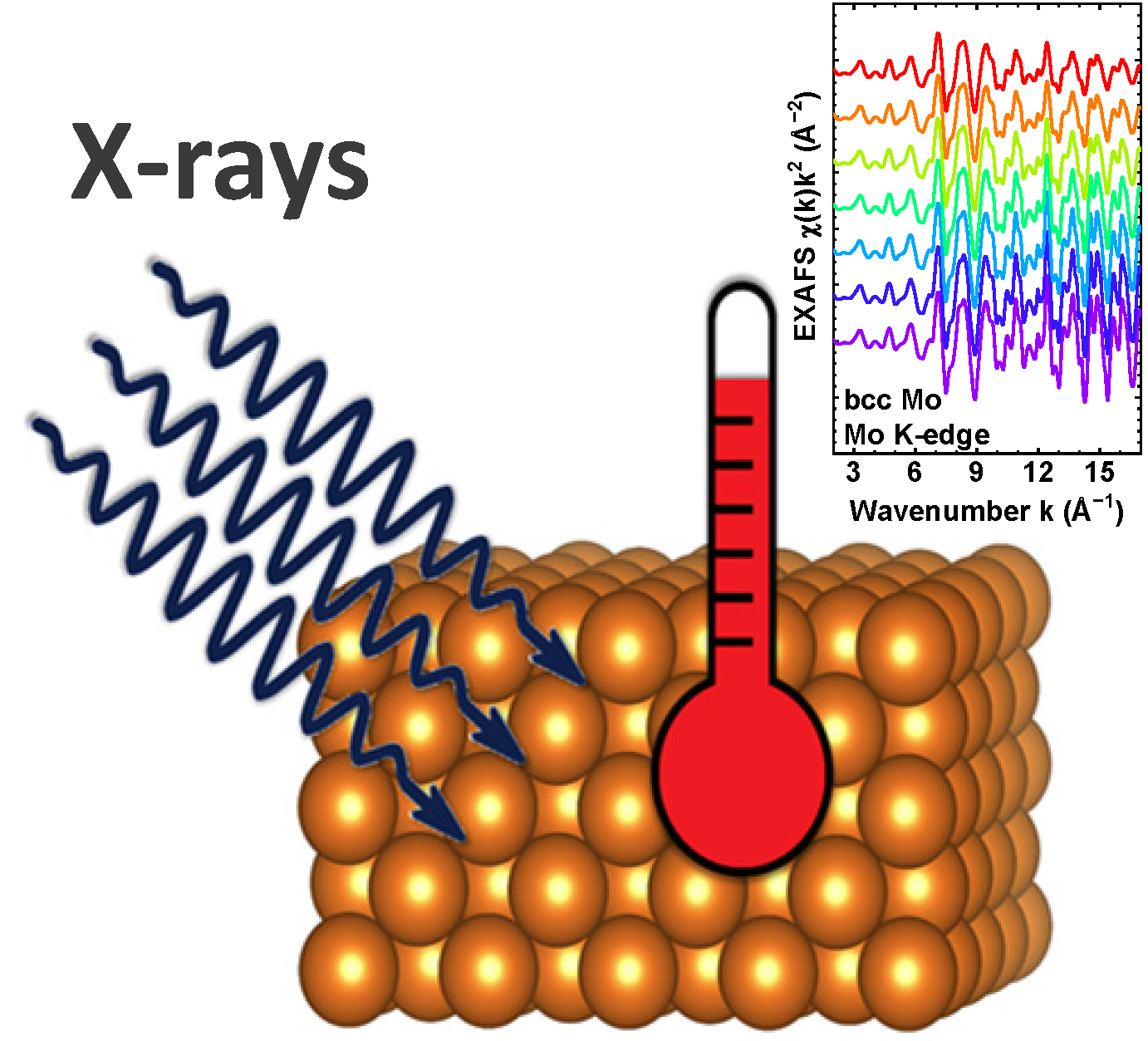}
  \medskip
  \caption*{Extended X-ray absorption fine structure (EXAFS) spectra are highly sensitive to thermal disorder. The temperature dependencies of the mean-square relative displacements are used to determine the local temperature of bcc (Cr, Mo, W) and fcc (Cu, Ag) metals from EXAFS spectra, based on the multiple-scattering formalism and the correlated Debye model. The pros and cons of the approach are discussed.}
\end{figure}

\end{document}